# Intelligent Risk Alarm for Asthma Patients using Artificial Neural Networks

Rawabi A. Aroud[1], Anas H. Blasi[2]

Computer Science Department[1]
Computer Information Systems Department[2]
Mutah University Al Karak, Jordan[1, 2]

Mohammed A. Alsuwaiket[3]

Computer Science and Engineering Technology Department
Hafar Batin University, Hafar Batin
Saudi Arabia

*Abstract*—**Asthma is a chronic disease of the airways of the lungs. It results in inflammation and narrowing of the respiratory passages; which prevents air flow into the airways and leads to frequent bouts of shortness of breath with wheezing accompanied by coughing and phlegm after exposure to inhalation of substances that provoke allergic reactions or irritation of the respiratory system. Data mining in healthcare system is very important in diagnosing and understanding data, so data mining aims to solve basic problems in diagnosing diseases due to the complexity of diagnosing asthma. Predicting chemicals in the atmosphere is very important and one of the most difficult problems since the last century. In this paper, the impact of chemicals on asthma patient will be presented and discussed. Sensor system called MQ5 will be used to examine the smoke and nitrogen content in the atmosphere. MQ5 will be inserted in a wristwatch that checks the smoke and nitrogen content in the patient's place, the system shall issue a warning alarm if this gas affects the person with asthma. It will be based on the Artificial Neural Networks (ANN) algorithm that has been built using data that containing a set of chemicals such as carbon monoxide, NMHC (GT) acid gas, C6H6 (GT) Gasoline, NOx (GT) Nitrogen Oxide, and NO2 (GT) Nitrogen Dioxide. The temperature and humidity will be also used as they can negatively affect asthma patient. Finally, the rating model was evaluated and achieved 99.58% classification accuracy.**

## I. INTRODUCTION

The development of human beings today has led to a heavy price. It is pollution that exists in our time, which increases continuously with every drop of fuel burned by human, and with declining air quality in urban areas, the risk of stroke, heart disease, lung cancer and acute and chronic respiratory diseases, including asthma is increasing [1]. Moreover, Air Pollution and Children's Health report, according to the World Health Organization in 2018, shows that 93% of children worldwide those under the age of fifteen breathe polluted air that puts their health and development at great risk. Estimates that around 600,000 children died in 2016 from acute respiratory infections caused by polluted air [2].

Asthma is a condition in the airways that occurs in the lungs. The muscles tighten around the airways, and excess swelling and irritation of the airways is called inflammation. In fact, it causes narrowing of the airways, coughing, wheezing, chest tightness, or trouble breathing. So, if asthma is left untreated, it may cause long-term lung function loss. Furthermore, when you are exposed to an asthma trigger, the air passages become more inflammatory or swollen than usual, making breathing more difficult or making illnesses worse [3].

The chemicals that have been studied in this paper are affecting the asthma patients negatively. For example, carbon monoxide is produced from partial oxidation (incomplete combustion of carbon) and organic compounds such as coal, this occurs when oxygen is scarce, or when the heat is very high. NMHC (GT) acid gas, especially in the natural gas field, is any gas mixture containing significant amounts of Hydrogen Sulfide (H2S) or carbon dioxide (CO2) or similar gases with an acidic character. C6H6 (GT) Benzene (or benzol) is liquid volatile color and one gasoline vehicles (fuel) and its highly flammable fumes are carcinogens and have a strong smell and jet. NOx (GT) Nitrogen Oxide is also known as dioxide nitrogen oxide or nitrogen monoxide, it is famous in the name of laughing gas for its stimulant effects when inhaled, it is a chemical compound with the chemical formula N2O, in the natural state it is a colorless gas, non-flammable, has a pleasant breath. NO2 (GT) Nitrogen Dioxide. Nitrogen dioxide is one of many nitrogen oxides, having the formula NO2 is a natural gas, brownish-red in color with a sharp pungent odor.

Other important factors have been considered in this paper are the temperature which can cause irritation or inflammation in the airways of an asthma patient, and Relative Humidity (RH) which is the biggest problem for asthma patients, it increases airway resistance to air flow, in addition to narrowing the bronchi, provoking coughing and an increase in mucus secretion, the secret of the bed bug recovering with high humidity is hidden to everyone. In addition to reducing vitamin.

Finally, the Absolute Humidity (AH) which may cause shortness of breath and respiratory diseases in healthy children. In addition, it could be the reason behind the emergence of asthma in the most sensitive and exposed individuals and also this may be due to mold, fungi, bacteria, dust insects, and even cockroaches.

This paper aims to predict the environmental chemicals such as chlorine gas, sulfur dioxide and smoke, which are the most gases affecting the asthma patient using Artificial Neural

Networks (ANN). The focus of this paper is on asthma patient. In general, asthma occurs when inhaling chemical vapors or gases. In 2015, the number of asthma patients exceeded 358 million compared to 183 million in 1990, causing 397,100 deaths in 2015 [4]. There is no effective treatment for asthma, but symptoms can be alleviated. It is necessary to follow a specific plan for the proactive management and control of symptoms. This plan includes reducing exposure to allergens and assessing the severity of symptoms and the use of medications [5].

Researchers recently created an artificial intelligence-based diagnostic algorithm by programming a GPU to act as a neural network, and by applying deep learning using a GPU. However, the team trained the neural network to identify and differentiate diseases. Finally, they showed a reliable result with high accuracy [6]. In this paper, a sensor system called MQ5 is used to examine the smoke and nitrogen content in the atmosphere. It is inserted in a wristwatch that checks the smoke and nitrogen content in the person's place, the system shall issue a warning if this gas affects the person with asthma. It will be based on the ANN algorithm.

The paper is organized as follows: Section 2 reviews the related work about predicting the risk of asthma symptoms using different data mining techniques, Section 3 describes the process followed to prepare the data including data understanding, selecting, transforming, and model building. Section 4 describes the interpretation and evaluation of the results. Finally, Section 5 discusses the conclusions and draws the future work.

## II. RELATED WORK

There are a large number of research papers were used data mining in asthma, some of them concentrate on predicting the risk of asthma symptoms' and others on understanding the relationship between allergens and asthma [7], including those who focused on smart diagnosis of asthma [8] and so on. However, the difference between this paper and other studies is that this paper is predicting the chemicals examined by a sensor called MQ5 to relieve the symptoms of asthma, so that a person with asthma should take precaution from where they are located. While other studies are focusing on the causes and symptoms of asthma and very limited who used the data mining for this matter.

In [9], a skilled diagnosis system for asthma and pulmonary embolism was developed and an algorithm to correctly distinguish between asthma and pulmonary embolism was developed as well. The researcher collected the data where he obtained 3657 records of the disease and the data were processed. Artificial neural networks algorithm was used to determine the need to operate the EDS system and perform confirmatory. Artificial Neural Networks algorithm (ANN) has a high 95% accuracy for asthma and pulmonary embolism samples, among the 1492 patients with respiratory disease, 1442 were classified correctly.

The authors [10] talked about the initial prediction of asthma away from the traditional method and with the help of Deep Neural Network (DNN) and Support Vector Machine (SVM). The aim of this study is to develop an algorithm and determine its effectiveness in the diagnosis of asthma. Data were obtained from Kendai University hospital. 566 patient records were collected. However, the search network was conducted on the basis of medium, after using both algorithms, the deep DNN obtained a high accuracy rate with 98% of the total prediction.

Another paper [11], the researcher sought to predict the risk of asthma attacks using machine learning approaches such as naïve Bayes (NB), Support Vector Machines (SVM), and Random Forests (RF). However, the study was conducted on 5 million records of infected patients and the goal was to reduce mortality so that it works to predict early in the risk that causes death. Logistic regression was used which was optimal in predicting the event. The data was validated using the Asthma Learning Health System (ALHS) and was developed to validate the asthma health system educational model. This work was carried out and with the support of asthma in the UK also an (ALHS) data set was established with funding from the National Council for Environmental Research (NCER).

Another paper focused on skilled diagnosis of asthma through machine learning algorithms [12], the k-NN and SVM algorithms were used, in addition to 169 people with asthma were tested and set of processes were used to implement the algorithm such as input organization, preprocessing, data tuning, and output evaluation. Tehran hospital was used to obtain the data, 250 records were taken, and data processing was done in two steps the first was the removal of incomplete data and the second step is to select the most important features that can be utilized in the algorithm. In the results, the researcher obtained the data through Canvas Orange and the implementation was in Python. Finally, SVM algorithm achieved the best results.

Another paper where authors introduced the development of the Lasso logistic regression model in 2015 [13] to predict asthma. In this research, the focus was on pediatric patients receiving medical care. The goal was to use administrative claims data for pediatric residents enrolled in Medicaid to train and test already deep in practice by comparing their predictive power. The Lasso logistic regression model served as a benchmark comparing the results of the deep learning model.

According to [14], the authors talked about machine learning was applied to the continuous biomarker so that the data provides an automatic respiratory novel for asthma in children using the Pediatric Asthma Guide (PAS) as a standard for clinical care. The ANN algorithm was applied to create an automatic respiratory score and validated by two approaches. However, ANN was compared with normal regression models and Poisson. Finally, results obtained an initial group of 186 patients and 128 patients met the inclusion criteria.

According to [15], the authors focused on predicting the disease of asthma using machine learning classification algorithms. Authors in this paper were used some machine learning algorithms such as SVM, ANN, k-NN and random forest algorithms. SVM algorithm have achieved 98% compared to other algorithms. MLP achieved 100% specificity compared to other algorithms. ANN achieved 100% sensitivity compared to other algorithms.

Due to the large proportion of people who affected by asthma in this world. Researchers have prepared studies to reduce its risks. Most papers have been working on developing a smart system that can distinguish between asthma and another disease, and others have predicted the initial diagnosis of asthma, also some of them who trained more than one algorithm to know which algorithms are the most accurate in predicting asthma and its symptoms.

This paper differs from other research in that it focuses on chemicals that have a significant impact on asthma patient in order to design a wrist watch can predict chemicals to issue a warning at the appropriate time that the patient's location is a danger to him/her to take an appropriate action to prevent the danger.

## III. Research Methodology

In order to get a better insight into the best predictability in chemicals a patient with asthma greatly helps to know the percentage of chemicals present everywhere. The dataset which collected for this study of the most common chemicals that appeared in one of the cities of Italy that contain gas Carbon monoxide, NMHC(GT), C6H6(GT), NOx (GT), and NO2(GT). These gases and other factors such as temperature and humidity would help to test and extract the model and know the accuracy that we can get from the ANN algorithm.

The critical step here as shown in Fig. 1 is a Knowledge Discovery in Databases (KDD) methodology which will be used as a methodology to manage all the processes that include data selection, preprocessing, data cleansing, building a data mining model and evaluating the results.

### A. Selection

Data were obtained from the UC Irvine Machine Learning Repository website [17] containing 9358 rows of decimal numbers for chemicals within an Italian city. Data recorded from March 2004 to February 2005. The data set properties are multivariate; time series and the attribute characteristics is real.

### B. Preprocessing

The data contained a set of empty and missing rows data was reorganized by disposing of empty and incomplete rows using Python programing language to convert numbers to be between 0 and 1. Data transformation through the alternative standardization is scaling features to lie between a given minimum and maximum value, often between zero and one, or so that the maximum absolute value of each feature is scaled to unit size. This can be achieved using MinMaxScaler or MaxAbsScaler in Python.

Table I shows a sample preprocessed data, where the target (output) is representing the scale of risk between 0 (lowest risk) and 1 (highest risk) for asthma patients, while the other eight attributes are the chemicals (inputs) of the proposed model.

### C. Data Mining

In the past ten years, Artificial Intelligence (AI) systems have been the best performing. Deep learning is actually a new name given to the AI approach and it has been called Artificial Neural Networks (NN), which started a long time ago more than 70 years ago. ANN were first proposed in 1944 by Warren McCullough and Walter Bates, researchers at the University of Chicago who moved to the Massachusetts Institute of Technology in 1952.

ANNs were a major area of research in both neuroscience and computer science until 1969. To do machine learning, where the computer learns to perform some tasks by analyzing training examples [18]. The structure and operation of the ANN can be described by the abstract model of the neural network of neurons, also called units or nodes. They can capture information from outside or from other neurons, transfer them to other neurons, or output them as a final result. There are positive and negative weights that represent an exciting or inhibiting effect. If the weight is zero, one neuron does not affect communication on the other hand. Neural networks can have a variety of different structures. These networks are also referred to as feedback networks or feedback networks [19][20]. In addition to a simple visualization mentioned in Fig. 2 to show how the inputs compared with each other.

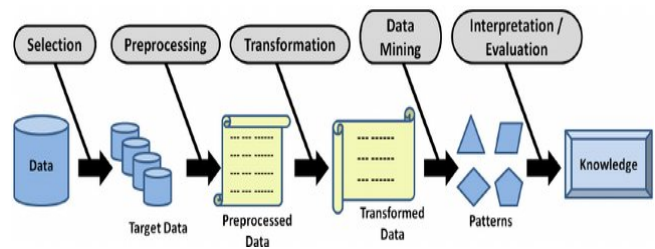

Fig. 1. Knowledge Database Discovery (KDD) Processes [16].

TABLE I. Sample of Preprocessed Data

| Carbon monoxide | NMHC (GT) | C6H6 (GT) | NOx (GT) | NO2 (GT) | T | RH | AH | Target |
|---|---|---|---|---|---|---|---|---|
| 0.66 | 0.79 | 0.79 | 0.60 | 0.58 | 0.6 | 0.6 | 0.8 | 0.66 |
| 0.84 | 0.47 | 0.47 | 0.93 | 0.76 | 0.4 | 0.7 | 0.5 | 0.65 |
| 0.78 | 0.72 | 0.72 | 0.80 | 0.51 | 0.4 | 0.8 | 0.5 | 0.65 |
| 0.9 | 0.35 | 0.35 | 0.93 | 0.65 | 0.3 | 0.8 | 0.4 | 0.65 |
| 0.85 | 0.35 | 0.35 | 0.92 | 0.67 | 0.3 | 0.8 | 0.5 | 0.64 |
| 0.64 | 0.79 | 0.79 | 0.55 | 0.51 | 0.6 | 0.7 | 0.8 | 0.64 |
| 0.96 | 0.35 | 0.35 | 0.91 | 0.56 | 0.3 | 0.8 | 0.4 | 0.64 |
| 0.66 | 0.79 | 0.79 | 0.60 | 0.58 | 0.6 | 0.6 | 0.8 | 0.66 |

ANNs represent a series of algorithms that seek to identify basic relationships in a set of data through a process that mimics the way the human brain works. It can adapt to changing inputs; therefore, the network generates the best possible result without having to redesign the output standards. In fact, the algorithm of the neural network in this paper was the best in predicting the chemical gases in the weather. ANN algorithm will be explained using Python programming language.

A graphical representation of the proposed ANN as mentioned in Fig. 2. However, the first set of 8 nodes is the inputs. The second set of 5 nodes is the hidden layer. The last set of three nodes is the output layer.

Fig. 3 shows how the inputs have compared to each other and how to represent the ratios of the chemicals that were recorded, in addition to represent the outputs in terms of risk with representations of the impact on the asthma patient. However, the high risk is in blue color, the medium risk is in orange color, and the low risk is in green color.

Python programming language is considered the most suitable and has been used in this paper because it is designed to be extendable with compiler code for proficiency, also many tools are available to facilitate Python integration and software code.

In this paper, a deep learning model has been applied using ANN algorithm to build the classification model. As mentioned previously in this paper, the ANN algorithm is a neural network of nutrition consisting of more than one hidden nonlinear layer. It is characterized by a combination of weight matrices, bias vectors, and a non-linear activation function [21]. Then, the ANN algorithm was built to construct the required model with combinations of testing and training the input layer.

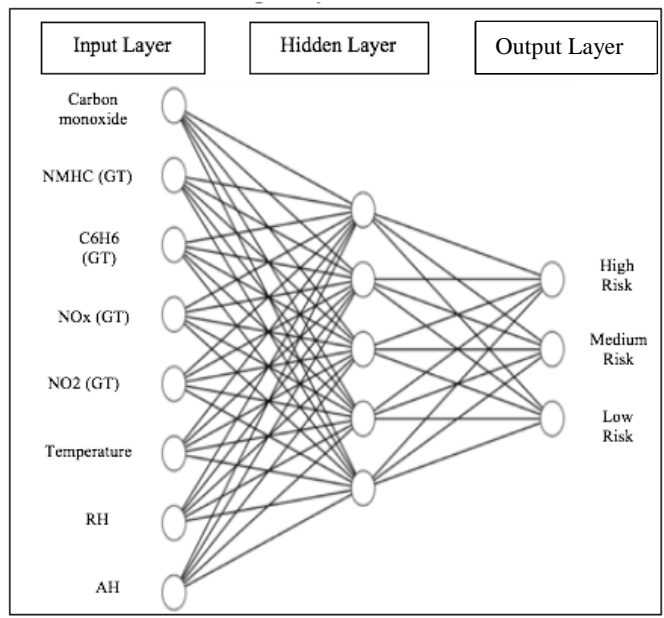

Fig. 2. Graphical Representation of the Proposed ANN.

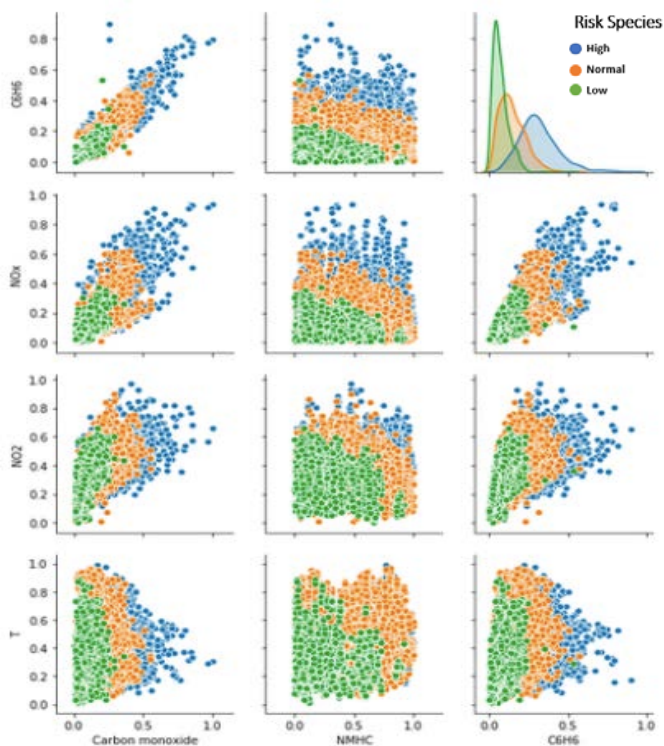

Fig. 3. Sample of Inputs Visualization with Comparison.

The proposed model has been used to predict the risk of these chemicals present in the region if they are high, medium or low risk for the asthma patients. The data have been divided into two sets; one is the training set with 70% of the data used to train the classifier for the prediction result. The other set is the test set with 30% of the data used to test the classifier. The proposed ANN has three layers, input, hidden and output layer. However, the Input layer has all the chemical gases on which the study was conducted with temperature and humidity. On other side, the output layer was made up of three values, each one indicating the types of risk, meaning that the values are focused on the type of risk, if it is high, medium or low risk. The weight matrix has been used to connect the inputs to the hidden layer of the proposed ANN. Each node in the input layer is connected to each node in the hidden layer. Weight values were randomly selected between -1 and 1. The Classification Accuracy (CA) that achieved from the proposed model is 99.58%.

## IV. RESULTS EVALUATION

In recent years, education has revolutionized science and knowledge in machine learning, especially seeing the computer. In this approach, the Artificial Neural Network (ANN) is trained, often in a supervised manner, using backpropagation due to an ANN and an error function, the method calculates the error function of the neural network weights. Huge quantities of specific training examples are needed, but the resulting classification accuracy is impressive, and sometimes it beats humans. The application is implemented in Python programming language.

The results obtained are in terms of percentage of accuracy. ANN gives high accuracy which is 99.58%. It was rated that the proportion of chemical gases if exceeded 0.6 out of 1, it will be considered high risk and can affect the asthma patient and in this case the patient will be warned to move from the area of risk. If the chemical gases are in between 0.3 and 0.6, it will be considered medium risk of the asthma patient and also in the case the patient will be notified of the level of risk, but if it was less than 0.3 out of 1, the proportion of chemical gases do not affect the asthma patient and will not warn the patient for any risk.

Table II shows the stratified cross-validation that seeks to ensure that each fold is representative of all strata of the data. Generally, this is done in a supervised way for classification and aims to ensure each class is approximately equally represented across each test fold (which are of course combined in a complementary way to form training folds), knowing that the weight values were randomly selected between -1 and 1 using Python.

Stratified cross validation in Table II consists of some important evaluation methods that assess the proposed model of ANN. Here the value of correctly classified instances (Classification Accuracy) is very high with 99.58, and the value of Mean Square Error (MSE) is very low with 0.0028. However, these results show that the proposed model has achieved very good results.

The intuition behind this relates to the bias of most classification algorithms in Table III detailed accuracy by class obtained after feature selection. They tend to weight each instance equally which means overrepresented classes get too much weight. Table III shows some evaluation methods for each class (High, Normal, Low).

However, these methods are True Positive rate (TP Rate) which represents the predicted instances as positive and are actually positive (higher better), False Positive (FP) which represents the predicted instances as positive and are actually negative (lower better), Precision which is the percentage of positive instances out of the total predicted positive instances (higher better), Recall which is the percentage of positive instances out of the total actual positive instances (higher better), F-measure is the harmonic mean of precision and recall. This takes the contribution of both, so higher the F1 score, the better. Finally, ROC Area stands for receiver operating characteristic and the graph is plotted against TPR and FPR for various threshold values. As TPR increases FPR also increases.

In general, all results are closely similar to each other and they have very competitive results.

TABLE II.    STRATIFIED CROSS-VALIDATION

| | |
|---|---|
| Correctly classified instances | 99.5822 |
| Incorrectly classified instances | 0.4178 |
| Kappa statistic | 0.9925 |
| Mean absolute error | 0.0028 |
| Root mean squared error | 0.0528 |
| Relative absolute error | 0.7407 |
| Root number of instances | 12.1721 |

TABLE III.    DETAILED EVALUATION METHODS BY CLASS

| TP Rate | FP Rate | Precision | Recall | F-measure | ROC Area | Class |
|---|---|---|---|---|---|---|
| 0.996 | 0.002 | 0.991 | 0.996 | 0.993 | 0.987 | High |
| 0.996 | 0.005 | 0.997 | 0.996 | 0.996 | 0.995 | Normal |
| 0.995 | 0.001 | 0.998 | 0.995 | 0.997 | 0.994 | Low |
| 0.996 | 0.003 | 0.996 | 0.996 | 0.996 | 0.993 | Weighted avg. |

TABLE IV.    CONFUSION MATRIX

| a | b | c | Classified as |
|---|---|---|---|
| 1358 | 6 | 0 | a= High |
| 13 | 4113 | 3 | b= Normal |
| 0 | 7 | 1441 | c= Low |

A confusion matrix as shown in Table IV is often used to describe the performance of a classification model (or "classifier") on a set of test data for which the true values are known. The confusion matrix itself is relatively simple to understand. Here, the row represents the predicted instances and the columns represents the actual instances for the dataset.

## V.    CONCLUSION AND FUTURE WORK

This study demonstrates that machine learning techniques such as ANNs were utilized to analyze parameters of simple vital signs and finite data. The potential impact of such an outcome is to improve and standardize data management to see the aggravation of acute asthma. This paper revealed several barriers to the integration of disparate data sources, it also processed and disposed of incomplete data. Further validation of the algorithm is imperative to improve data integrity, and to improve and expand the contribution features. Because asthma in children is the most prevalent chronic childhood disease in the future, this study has endeavor to design a wristband that contains a sensor MQ5 that examines chemicals in the weather and when the danger increases, it sends an alert message to the patient concerned alert him/her has been exposed to high pollution.

In future work, a different data set will be applied from different regions of the world and different settings of hidden layers will be tested in ANN as well. In addition to use other machine learning algorithms such as decision tree DT [22] or/and fuzzy logic [23], then compare the results with ANNs results.